\documentclass[aps,prl,twocolumn,superscriptaddress,letterpaper]{revtex4-2}

\usepackage{graphicx}
\usepackage{bm}
\usepackage{amssymb}
\usepackage{color}
\usepackage{amsmath}
\usepackage{ulem}
\usepackage[colorlinks=true,linkcolor=blue,anchorcolor=black,citecolor=blue,filecolor=black,menucolor=black,runcolor=black,urlcolor=blue]{hyperref}

\begin{document}

\title{Universal valley filtering via uniform dissipation and velocity contrast}

\author{Sijie Yue}
\thanks{These authors contributed equally to this work}
\affiliation{National Laboratory of Solid State Microstructures, School of Physics, Jiangsu Physical Science Research Center, and Collaborative Innovation Center of Advanced Microstructures, Nanjing University, Nanjing 210093, China}

\author{Wentao Xie}
\thanks{These authors contributed equally to this work}
\affiliation{Department of Physics, The Chinese University of Hong Kong, Shatin, Hong Kong SAR, China}

\author{Kai Shao}
\thanks{These authors contributed equally to this work}
\affiliation{National Laboratory of Solid State Microstructures, School of Physics, Jiangsu Physical Science Research Center, and Collaborative Innovation Center of Advanced Microstructures, Nanjing University, Nanjing 210093, China}
\affiliation{School of Physics and Astronomy, Yunnan Key Laboratory for Quantum Information, Yunnan University, Kunming 650091, People’s Republic of China}

\author{Hong-yu Zou}
\affiliation{Research Center of Fluid Machinery Engineering and Technology, School of Physics and Electronic Engineering, Jiangsu University, Zhenjiang 212013, China}

\author{Bingbing Wang}
\affiliation{Department of Physics, The Chinese University of Hong Kong, Shatin, Hong Kong SAR, China}

\author{Hong-xiang Sun}
\email{jsdxshx@ujs.edu.cn}
\affiliation{Research Center of Fluid Machinery Engineering and Technology, School of Physics and Electronic Engineering, Jiangsu University, Zhenjiang 212013, China}
\affiliation{State Key Laboratory of Acoustics, Institute of Acoustics, Chinese Academy of Sciences, Beijing 100190, China}

\author{Y. X. Zhao}
\affiliation{Department of Physics and HKU-UCAS Joint Institute for Theoretical and Computational Physics at Hong Kong, The University of Hong Kong, Pokfulam Road, Hong Kong, China}
\affiliation{HK Institute of Quantum Science \& Technology, The University of Hong Kong, Pokfulam Road, Hong Kong, China}

\author{Wei Chen}
\email{pchenweis@gmail.com}
\affiliation{National Laboratory of Solid State Microstructures, School of Physics, Jiangsu Physical Science Research Center, and Collaborative Innovation Center of Advanced Microstructures, Nanjing University, Nanjing 210093, China}

\author{Haoran Xue}
\email{haoranxue@cuhk.edu.hk}
\affiliation{Department of Physics, The Chinese University of Hong Kong, Shatin, Hong Kong SAR, China}
\affiliation{State Key Laboratory of Quantum Information Technologies and Materials, The Chinese University of Hong Kong, Shatin, Hong Kong SAR, China}

\begin{abstract}
Valley, as a ubiquitous degree of freedom in lattices, has found wide applications in both electronic and classical-wave devices in recent years. However, achieving valley-polarized states, a prerequisite for valley-based operations, still remains challenging. Here, we propose and experimentally demonstrate a universal non-Hermitian mechanism for valley filtering using only uniform background dissipation, which creates a propagation length contrast between valleys through their intrinsic group velocity differences. We implement this concept in an acoustic crystal, observing switchable and robust valley polarization of sound through large-scale field mapping. Remarkably, our approach is solely based on uniform loss, without the need for any special lattice structures, tailored excitations, or external fields. We further provide designs of our non-Hermitian valley filter on photonic and electronic platforms. Our results offer a simple and effective solution to valley-polarized state generation and may advance the development of novel valley-based devices in both classical and quantum regimes.
\end{abstract}

\maketitle
\textit{Introduction.}---Valley, a local energy extremum in the band structure, has emerged as a fundamental degree of freedom to process information and manipulate energy flow in crystalline systems, including both solid-state materials~\cite{rycerz2007valley, xiao2007valley, xiao2012coupled, zeng2012valley, mak2012control, cao2012valley, mak2014valley, gorbachev2014detecting} and artificial crystals~\cite{lu2016valley, ma2016all, dong2017valley, lu2017observation, wu2017direct, gao2018topologically, noh2018observation}. Through several decades of development, various valley-based applications across multiple disciplines have been identified, such as valleytronics as a supplement for spintronics~\cite{schaibley2016valleytronics, vitale2018valleytronics} and valley photonics for next-generation photonic technologies~\cite{xue2021topological, liu2021valley}. Although the valley degree of freedom is ubiquitous, generating valley-polarized states, which are essential for valley-based applications, is still a crucial problem to be fully resolved.

Current methods to achieve full valley-polarized states can be classified into two categories: the selective excitation approach and the valley filtering approach, both of which are model-specific and involve a certain level of additional complexity to the system. For example, to selectively excite the modes in one valley, one usually needs a carefully designed excitation, such as a polarized laser for an electronic system~\cite{zeng2012valley, mak2012control, cao2012valley} and a phased array for a photonic system~\cite{chen2018tunable}. To achieve a valley filter, similarly, extra efforts cannot be avoided, which involve the need for external fields~\cite{mak2014valley, gorbachev2014detecting} or the engineering of sample geometries~\cite{rycerz2007valley, li2018valley, wang2022observation}. Therefore, a simple and universal route towards valley-polarized states is highly demanded.

Recent advances in nonconservative systems open new possibilities for valley filtering via non-Hermitian mechanisms~\cite{el2018non, ashida2020non, bergholtz2021exceptional, ding2022non}. Non-Hermiticity, which causes the states to decay or grow, naturally aligns with the functionality of valley filters that need to get rid of the modes in one of the valleys. A straightforward thinking is to make the lifetimes of the propagating modes towards the same direction different for the two valleys. Indeed, this can be done by using a moir\'{e} structure~\cite{shao2024non} or by incorporating long-range non-Hermitian hoppings~\cite{xie2025non}. However, these non-Hermitian valley filters again require high complexity in sample design and fabrication, hindering practical applications.

In this work, we propose a non-Hermitian valley filter using solely a uniform background dissipation. The key insight in our construction is that, instead of building a lifetime contrast between the two valleys, we alternatively opt for a contrast in the group velocity, which can also make the propagation lengths of the modes at the two valleys unequal. In our design, the group velocity contrast, which needs not be large as we will demonstrate later, is naturally present in most valley systems. The other key factor, i.e., the background dissipation, is also naturally present in all wave systems and can be easily implemented in electronic systems with an extra layer serving as an environment. Hence, our design is platform-free and can bypass all previous complexities in achieving a valley filter. 

We experimentally demonstrate the idea using an acoustic crystal through direct phase-resolved field mapping and rigorous valley polarization analysis. The constructed valley filter is able to generate fully valley-polarized states with switchable valley polarizations, and has the advantages of superior efficiency (i.e., the valley polarization grows fast with filter length) and high robustness (i.e., the mechanism works under arbitrary excitation positions and local defects).  In addition, we provide two extra designs in Supplemental Material (see Sec.~S13 and Sec.~S14 in~\cite{SM}) using the same principle, for valley filters in photonic and electronic platforms, respectively, to show the universality of our approach.

\nocite{Ando_1991_PRB,Khomyakov2005PRB,yang2020terahertz,shalaev2019robust,Kužel,datta1997electronic,Nardelli1999_PRB}

\begin{figure}[h]
	\centering
	\includegraphics[width=\columnwidth]{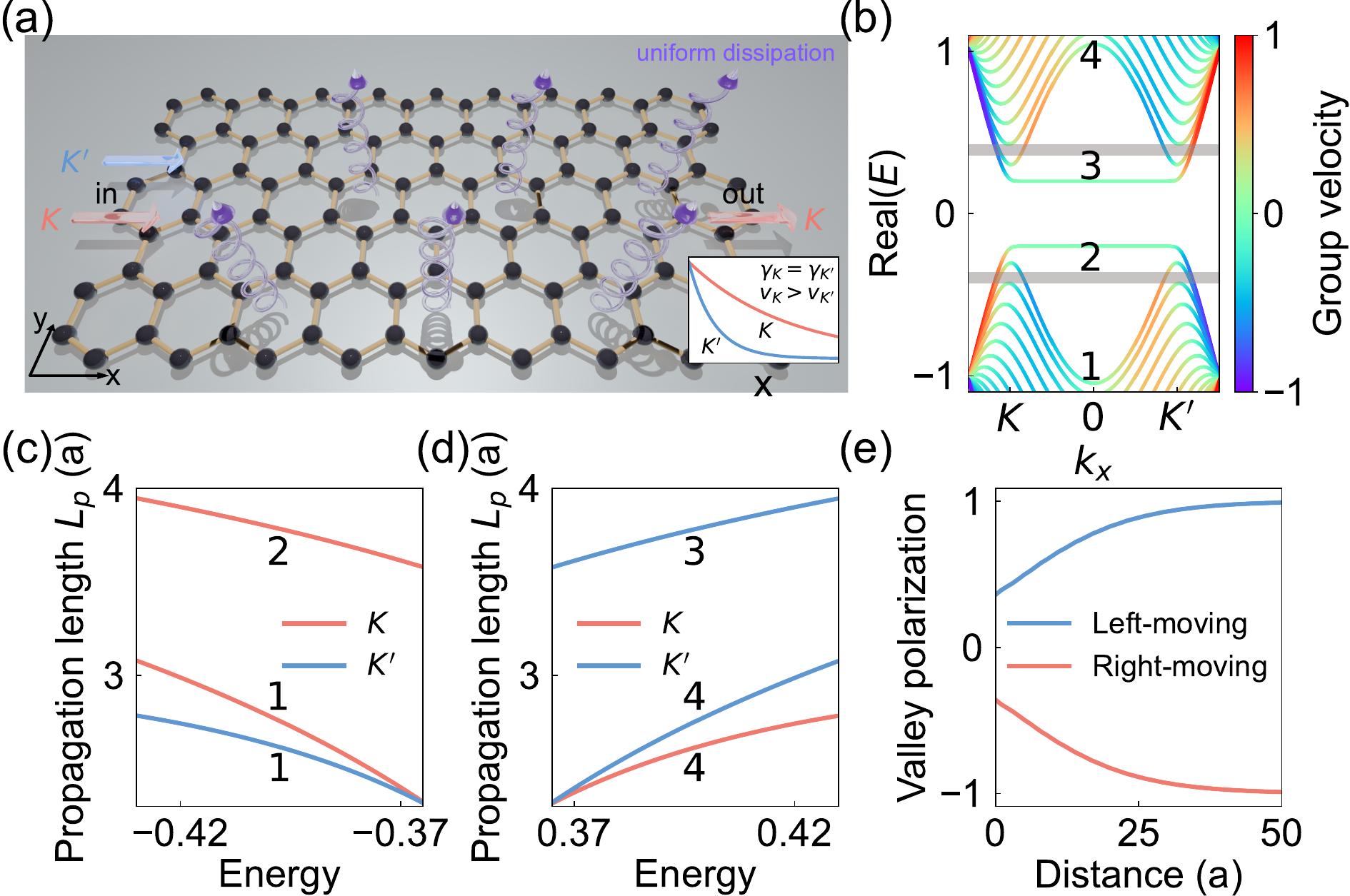}
	\caption{General construction of the non-Hermitian valley filter. (a)~Schematic of a non-Hermitian valley filter on a graphene lattice with a uniform background dissipation. The inset illustrates the unequal decay rates of the modes at the two valleys due to the difference in their group velocity. (b)~Band diagram for the model in (a) under a slab geometry with 20 unit cells along the $y$ direction. The color represents the group velocity. (c)-(d)~Propagation lengths of the right-moving modes near the valleys inside the energy windows indicated by the grey boxes in (b). (e)~Calculated valley polarizations against propagation distance for left-moving (blue curve) and right-moving (red curve) modes at $E=0.4$. In all calculations, we set the lattice constant to be unity, $t=1$, $m=0.2$, and $\gamma=0.1$.}
	\label{fig1}
\end{figure}

\textit{General tight-binding model.}---To illustrate our idea, we first consider a simple graphene model given by
\begin{equation}
\begin{split}
    H=&-t\sum_{<i,j>}\left(a_i^{\dagger}b_j+h.c.\right)+m\left(\sum_{i\in A}a_i^{\dagger}a_i-\sum_{i\in B}b_i^{\dagger}b_i\right)\\
    &-i\gamma\left(\sum_{i\in A}a_i^{\dagger}a_i+\sum_{i\in B}b_i^{\dagger}b_i\right),
\end{split}
\label{H}
\end{equation}
where $a_{i}^{\dagger}(a_{i})$ is the creation (annihilation) operator at the $i$-th site, $t=1$ is the nearest-neighbor coupling strength, $m$ is an on-site energy detuning, and $\gamma$ represents a uniform dissipation [Fig.~\ref{fig1}(a)]. In practice, $\gamma$ can be a result of intrinsic absorption or leakage to the environment, as we will demonstrate later. Figure~\ref{fig1}(b) plots the band structure under a slab geometry, which resembles that of a conventional gapped graphene~\cite{yao2009edge}, with the only difference being an overall shift of eigenenergies along the imaginary axis due to the uniform background dissipation $i\gamma$. So far, the effect of the dissipation seems trivial as it induces the same temporal decay for all eigenenergies and does not modify the mode profiles at all.

\begin{figure*}
	\centering
	\includegraphics[width=\textwidth]{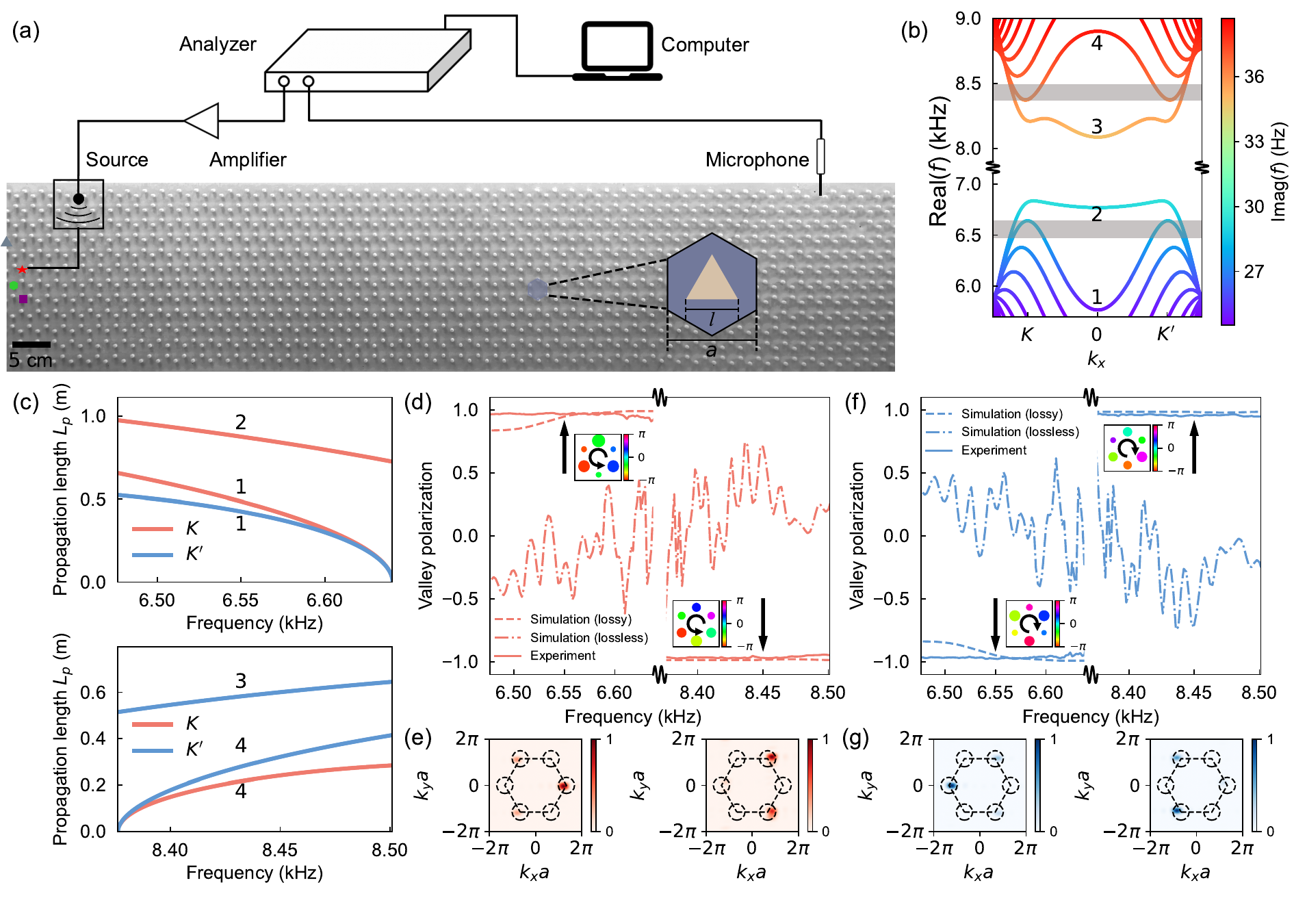}
	\caption{Non-Hermitian valley filter in an acoustic crystal. (a)~Schematic of the experimental setup, consisting of the fabricated sample, a speaker, a microphone, and a signal generation and data acquisition system. The inset shows the structural detail of the unit cell. (b)~Band diagram for the acoustic crystal under a slab geometry with 13 unit cells along the $y$ direction. The color represents the imaginary part of the eigenfrequency. (c)~Propagation lengths of the right-moving modes near the valleys inside the frequency windows indicated by the grey boxes in (b). (d)~Measured (solid curves) and simulated (dashed curves) valley polarizations under an excitation at the left side of the sample [indicated by the red star in (a)]. The valley polarization is calculated using the fields inside a region whose center is 1846.25 mm away from the source. The insets show the intensity and phase distributions inside a small area in the measured region. (e)~Measured Fourier spectra at the frequencies indicated by the black arrows in (d). The hexagon and the circle denote the first Brillouin zone and the area to evaluate the weights of the two valleys, respectively. (f)-(g)~The same as (d) and (e) but for a right-side excitation.}
	\label{fig2}
\end{figure*}

However, for a valley filter device, the relevant quantity is the propagation length rather than the lifetime, which characterizes the spatial decay of a mode and is given by
\begin{equation}
L_p=|v_g| \tau
,\end{equation}
where $v_g$ is the group velocity and $\tau$ is the lifetime defined by the inverse of the imaginary part of the eigenenergy. The propagation length corresponds to the distance a wave can travel before its amplitude decays to $1/e$ of its initial value. Importantly, this quantity is jointly determined by the lifetime and the group velocity, the latter of which is generally different for modes at opposite valleys, as indicated by the colors in Fig.~\ref{fig1}(b). We plot in Figs.~\ref{fig1}(c)-(d) the propagation lengths of the modes near the valleys (in the energy windows from -0.43 to -0.365 and from 0.365 to 0.43) and indeed find that there is a propagation length contrast for both lower and upper bands. This indicates that, after some propagation distances, one of the valleys will dominate, leading to a non-Hermitian valley filter. 

The analysis of the dispersion is verified through large-scale transport calculations using the scattering matrix method (see Sec.~S1 in~\cite{SM}). We use the valley polarization to characterize the valley filter performance, which is defined as
\begin{equation}
    P=\frac{W_K-W_{K^{\prime}}}{W_K+W_{K^{\prime}}},\label{pol}
\end{equation} 
where $W_K(W_{K^{\prime}})$ is the weight of $K$($K'$) valley modes in the transmitted wave. In the tight-binding calculations, it is evaluated through the valley-resolved conductance (see Sec.~S1 in~\cite{SM}). When $P=1(-1)$, the states are fully $K$($K'$)-polarized. The results show that the valley polarization approaches $\pm1$ after propagation of around 50 lattice constants, and the filtered valleys can be flipped by reversing the propagation direction [Fig.~\ref{fig1}(e)] as a consequence of the time-reversal symmetry (see Sec.~S2 in~\cite{SM}).

\textit{Non-Hermitian valley filter in an acoustic crystal.}---We demonstrate such a non-Hermitian valley filter in an acoustic crystal as shown in Fig.~\ref{fig2}(a), which consists of triangular scatterers arranged in a triangular lattice. This design is popular among classical wave platforms and has demonstrated excellent potential in various applications~\cite{lu2016valley, ma2016all, lu2017observation, wu2017direct, gao2018topologically, yan2018chip, shalaev2019robust, zeng2020electrically, yang2020terahertz, wang2024chip}. At the Brillouin zone corners, two Dirac cones emerge, whose masses depend solely on the orientation of the scatterers. Here, we make the scatterers point upwards, which breaks the reflection symmetry about the $x$ axis, to obtain two massive Dirac cones as normally adopted in previous studies. However, we note that the orientation of the scatterers is not crucial to achieve valley filtering (see Sec.~S3 in~\cite{SM}). 

To construct a valley filter device, we consider a strip geometry with periodic (hard) boundaries in the $x$ ($y$) direction (see Sec.~S4 and Sec.~S5 in~\cite{SM}), and the sound wave can be injected through either the left or right side of the system using a point-like source (see Sec.~S6 in~\cite{SM}). Throughout this work, we focus on the frequency ranges from 6476 Hz to 6638 Hz and from 8375 Hz to 8500 Hz, where the lower and upper bulk modes are well-defined valley modes located within $0.175\pi/a$ from the valley centers [Fig.~\ref{fig2}(b)]. The working frequency range can be further tuned by scaling the sample size.

At this stage, the two key factors to achieve the non-Hermitian valley filter are already fulfilled. Firstly, a uniform background loss naturally occurs due to the sound absorption in the air, leading to a momentum-independent imaginary part in the eigenfrequency (see Sec.~S7 in~\cite{SM}). In the numerical simulations, this loss is modeled by a nonzero imaginary part in the sound speed (see Sec.~S8 in~\cite{SM}). Secondly, the group velocities of the modes at opposite valleys are generally different, resulting in contrasting propagation lengths in the two valleys, as plotted in Fig.~\ref{fig2}(c).

To test the performance of the valley filter, we fabricate a large sample with 100 unit cells along the $x$ direction. We first place the speaker at the position of the red star in Fig.~\ref{fig2}(a) and measure the sound pressure in a region containing 75 unit cells along the $x$ direction. Figure~\ref{fig2}(d) plots the measured valley polarization [with $W_{K/K'}$ obtained by integrating the Fourier intensities inside a circular region of radius $0.35\pi/a$; see Fig.~\ref{fig2}(e)] after a long-distance propagation, which approaches 1 and -1 for the lower and upper bands, respectively, suggesting an excellent valley filter behavior. Here, the opposite valley polarization is caused by the opposite propagation length contrast for the lower and upper bands, as can be seen from Fig.~\ref{fig2}(c). Interestingly, we can also turn off the background loss numerically by setting the sound speed as a real number (experimentally, however, the loss is always present). Under this setting, the valley filter does not work anymore, as revealed by the oscillating valley polarization curve in Fig.~\ref{fig2}(d). This suggests that the background loss indeed plays a crucial role. 

The valley filter behavior can also be directly judged from the real-space fields. Valley materials are known to host vortices with valley-dependent chirality due to the broken inversion symmetry~\cite{xiao2007valley}. In our experiment, this is reflected by the phase winding in the measured fields. Here, the measurement points form a honeycomb structure [Fig.~\ref{fig2}(a)]. The valley modes are mainly contained in one of the sublattices and form valley-dependent phase windings for the three nearest sites of the same sublattice due to the Bloch phase~\cite{yao2008valley}. The insets to Fig.~\ref{fig2}(d) show the measured amplitudes and phases of sound in a small area at 6550 Hz (8450 Hz), which exhibit a clear anticlockwise phase winding that corresponds to the $K$($K'$) valley modes in the lower (upper) band (see Fig.~S6 in~\cite{SM}).

\begin{figure}
	\centering
	\includegraphics[width=0.9\columnwidth]{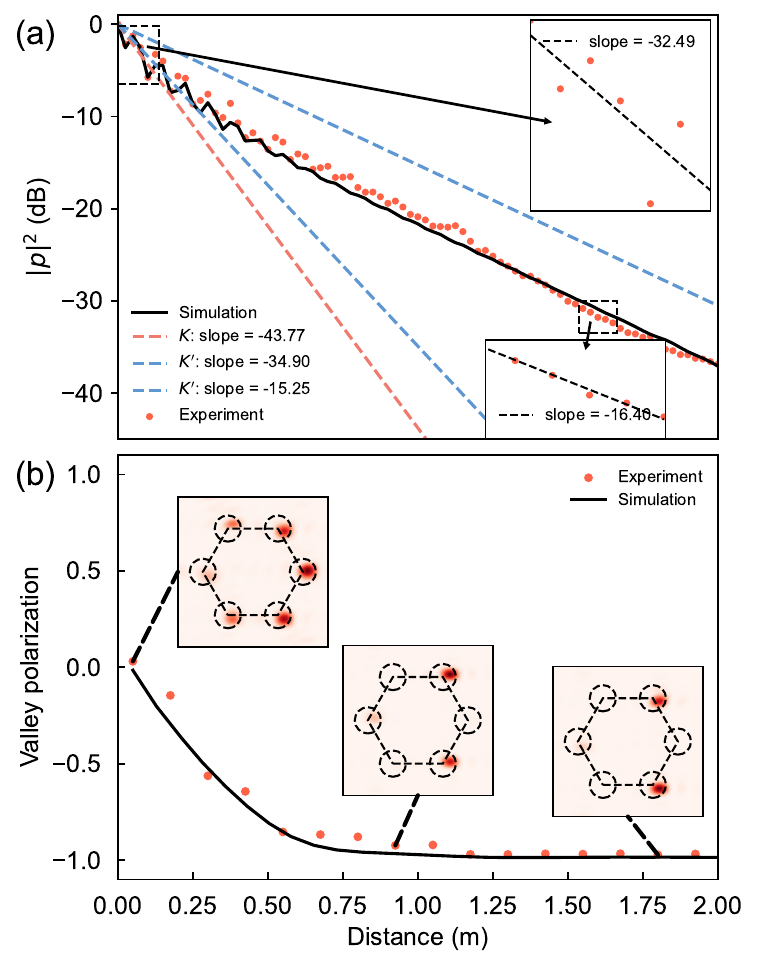}
	\caption{Analysis of the acoustic non-Hermitian valley filter. (a)~Plots of the measured (red dots) and simulated (black curve) acoustic intensities against propagation distances at 8420 Hz. The dashed lines represent the decays of the three eigenmodes at the same frequency. The insets show the enlarged plots of the measured data at the early and later stages of the propagation, with the dashed lines showing the linear fits. (b)~Plots of the measured (red dots) and simulated (black curve) valley polarizations against propagation distances at 8420 Hz. The insets display the measured Fourier spectra at three selected propagation distances.}
	\label{fig3}
\end{figure}

\begin{figure}
	\centering
	\includegraphics[width=\columnwidth]{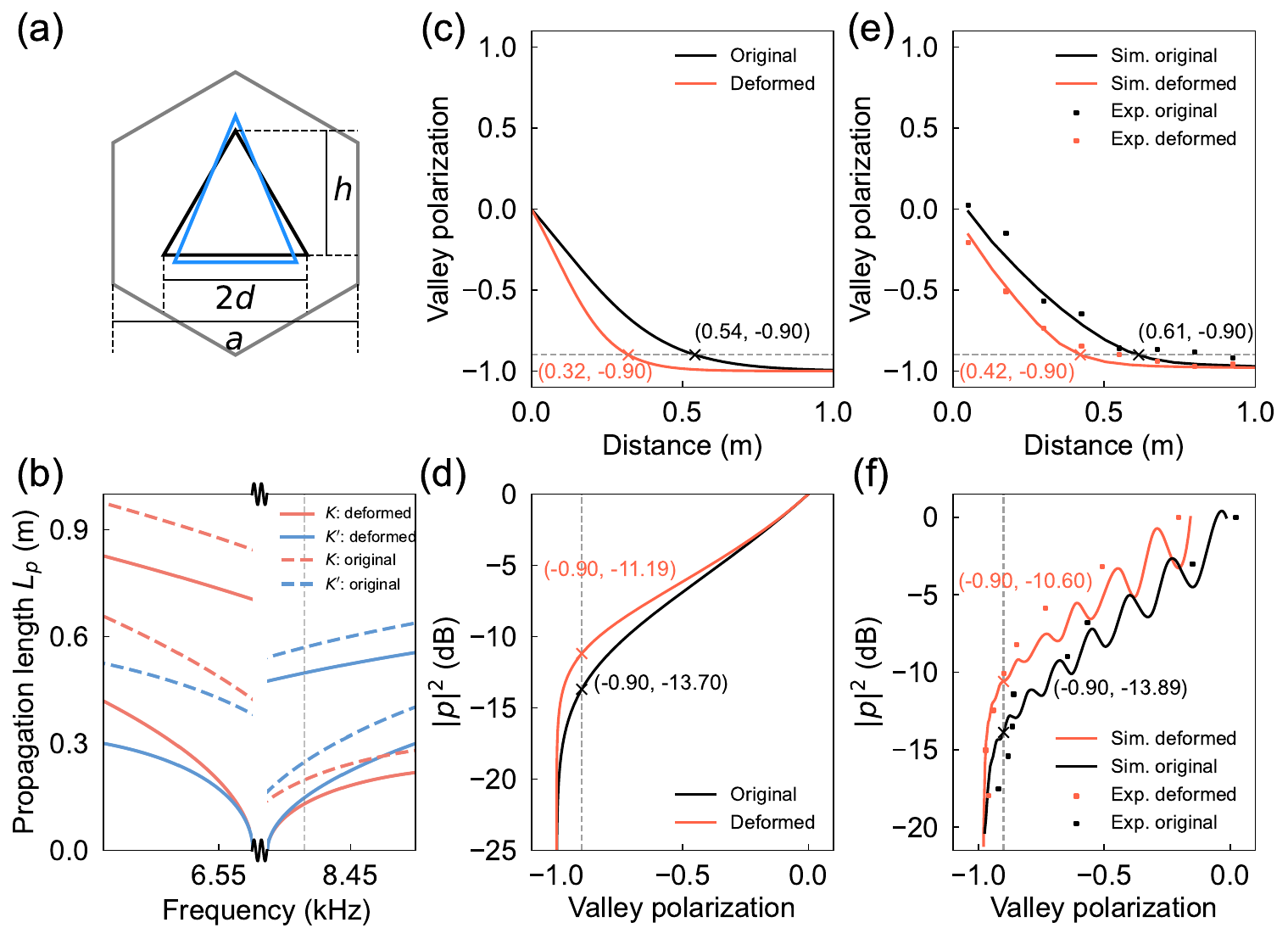}
	\caption{Enhancing the filtering performance. (a) Schematic of scatterer deformation. The black (blue) triangle denotes the original (deformed) scatterer of height $h$ and base length 2$d$. (b) Propagation lengths of the right-moving valley modes before (dashed curves) and after  (solid curves) the scatterer deformation. (c) Plot of theoretical valley polarization against propagation distance. (d) Plot of theoretical intensity against valley polarization. (e) Plots of simulated (solid curves) and measured (dots) valley polarization against propagation distance. (f) Plot of simulated   (solid curves) and measured (dots) sound intensity against valley polarization. In (c)-(f), black (red) curves correspond to data for the original (deformed) sample at 8420 Hz, and the highlighted data points correspond to the cases when -0.9  valley polarizations are achieved.}
	\label{fig4}
\end{figure}

Practically, one would expect that a valley filter can generate valley-polarized states for both valleys at a given frequency (or energy). To realize this functionality, we simply need to relocate the excitation from the left to the right side. Due to the time-reversal symmetry, two modes with the same frequency but opposite wavevectors must have their group velocities equal in amplitude but opposite in sign. Consequently, if the right-moving modes in the $K$ valley have a larger propagation length, their time-reversed left-moving modes in the $K'$ valley will also have a larger propagation length (see Sec.~S2 in~\cite{SM}). Hence, after the excitation is moved to the right side, the filtered valley is also flipped [see Figs.~\ref{fig2}(f)-(g)].

Next, we proceed to a detailed quantitative analysis to gain a deeper understanding of the working principle. To this end, we plot in Fig.~\ref{fig3}(a) the sound decay over propagation at 8420 Hz, which is obtained by integrating the sound intensity along the $y$ direction at each fixed $x$. As references, we also plot in dashed lines the decays of the three right-moving eigenmodes at the same frequency, calculated from their group velocities and lifetimes. Since our excitation, in general, will excite all three modes, the experimental decay curve does not follow any of the dashed lines. Instead, it lies in between them, reflecting that the excitation contains all three modes. Furthermore, as the three modes carry different propagation lengths, the experimental field does not host a constant decay against propagation distance. In fact, by taking a closer look, we can find that the decay, measured by the slope of the curve, aligns with the decay of $K$ and $K'$ valley modes at short and long propagation distances, respectively [see the insets of Fig.~\ref{fig3}(a)]. This reveals that, due to the propagation length contrast, the $K$ valley modes are washed out during the early stage propagation, while the $K'$ valley modes dominate the long-distance decay behavior [Fig.~\ref{fig3}(b)].

The above analysis suggests that our non-Hermitian valley filter is insensitive to the initial weights of the valleys. Even if the unwanted valley is dominantly excited, its weight will decrease quickly and finally approach zero during propagation due to its shorter propagation length (see Sec.~S9 in~\cite{SM}). This property also endows the valley filter with robustness against sample defects. This is because, while a defect can cause inter-valley scattering, the induced change in valley polarization will soon be compensated by the non-Hermitian filtering process during propagation (see Sec.~S10 in~\cite{SM}). It is also worth pointing out that, despite the signal decay induced by dissipation, the signal-to-noise ratio remains practically acceptable (estimated to be around 28.8 dB when the valley polarization reaches -0.9), due to the fact that only background dissipation is present without involving any additional loss. Furthermore, the signal attenuation can be avoided by using gain (see End Matter).

\textit{Enhancing the filtering performance.}---Finally, we demonstrate that the performance of the valley filter can be further improved through a simple structural deformation. To reduce the sample size and enhance the sound intensity when a high valley polarization value (e.g. $\pm0.9$) is achieved, the group velocity needs to be tuned. For this purpose, we deform the original equilateral triangle to an isosceles one, as illustrated in Fig.~\ref{fig4}(a). During this process, the ratio $h/d$ ($h$ is the height and $2d$ the base length) changes from $\sqrt{3}$ to 2.4, while the area of the triangle ($hd=92.932$ mm$^2$) remains unchanged to maintain the working frequency. After the deformation, the propagation lengths of the modes experience noticeable changes [Fig.~\ref{fig4}(b)].

To theoretically analyze the filtering performance, we assume an unpolarized initial state (with the weights of the $K$ valley mode and each of the $K'$ modes being 50\% and 25\%, respectively) and evaluate its propagation in the original and deformed samples using the propagation length data at 8420 Hz. We find that the deformed sample outperforms the original one in the sense that it achieves a highly polarized state faster [Fig.~\ref{fig4}(c)] with higher signal strength [Fig.~\ref{fig4}(d)]. This finding is verified through simulation and measurement of a new sample consisting of the deformed scatterers [Figs.~\ref{fig4}(e)-(f)]. Interestingly, using a uniaxial strain, a similar effect is also observed in the graphene model (see Sec.~S11 in~\cite{SM}). In addition to group velocity engineering, we note that increasing the dissipation can also make the valley polarization converge faster (see Sec.~S12 in~\cite{SM}).

\textit{Conclusions.}---We have proposed a general approach to prepare fully valley-polarized states using a simple uniform background dissipation and experimentally demonstrated our proposal in an acoustic system. Our acoustic device exhibits superior valley filtering behaviors, including switchable valley polarization, fast filtering speed, and easy excitation, which pave the way for further construction of compact and multifunctional valley devices. While demonstrated in acoustics, our method can be straightforwardly applied to other wave systems (e.g., photonics) as long as a background dissipation is present (see Sec.~S13 in~\cite{SM}). Extending to electronic materials, we propose a generalization of our method by using a second layer as a source of non-Hermiticity (see Sec.~S14 in~\cite{SM}). 

An inspiration from our results is that, while engineering complex non-Hermitian textures results in fruitful physics, systems with uniform losses can also exhibit intriguing non-Hermitian phenomena despite the seemingly trivial effects of uniform losses on the eigenspace. The key is that some quantities, like the propagation length in this study, are jointly controlled by Hermitian and non-Hermitian factors. Considering the fact that systems with uniform losses are much easier to obtain, as almost all classical wave systems naturally fall into this category, discovering new physics and phenomena in these systems would greatly facilitate the physical relevance and potential applications of non-Hermitian physics across broad disciplines.

\begin{acknowledgments}
\textit{Acknowledgments.}---We are grateful to Zhesen Yang for helpful discussions. This work was supported by the National Natural Science Foundation of China under Grants No.~12222406 (W.C.), 62401491 (H.X.) and 12274183 (H.S.), the National Key Research and Development Program of China under Grant No.~2025YFA1412300 (H.X.), the Natural Science Foundation of Jiangsu Province under Grants No.~BK20250008 (W.C.), BK20253009 (W.C.), and BK20233001 (W.C.), the Research Grants Council of Hong Kong SAR, China, under Grants No.~24304825 (H.X.) and 17301224 (Y.X.Z.), the Jiangsu Qing Lan Project~(H.S.), the Guangdong Provincial Quantum Science Strategic Initiative under Grant No.~GDZX2501012~(H.X.), and the Chinese University of Hong Kong under Grants No.~4053729 (H.X.) and 4053794 (H.X.).
\end{acknowledgments}

\textit{Data availability.}---The data that support the findings of this study are openly available~\cite{yue_2026_21289325}.

\bibliography{references-main}

\bigskip
\clearpage

\section*{End Matter}

\begin{figure}[h]
	\centering
	\includegraphics[width=\columnwidth]{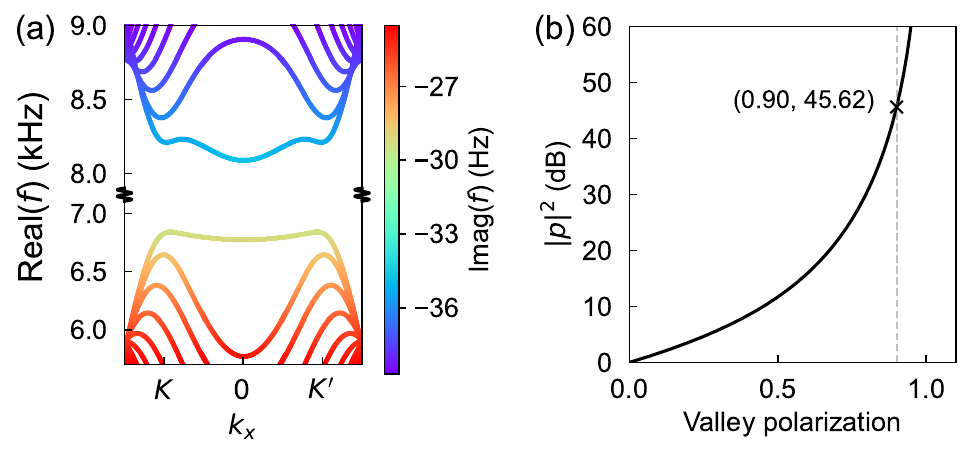}
	\caption{Acoustic valley filter via uniform gain. (a)~Band diagram for the acoustic crystal under a slab geometry with 13 unit cells along the $y$ direction. The color represents the imaginary part of the eigenfrequency. (b)~Plot of theoretical intensity against valley polarization at 8420 Hz. The vertical dashed line indicates the position at which the valley polarization reaches 0.9.}
	\label{fig5}
\end{figure}

\textit{Valley filtering via uniform gain.}---In the main text, our valley-filtering mechanism is implemented by using uniform dissipation. This implementation inevitably causes signal attenuation, leading to a reduced signal-to-noise ratio. Here we show that the working principle of the valley filter remains unchanged when the uniform dissipation is switched to uniform gain. For this purpose, we flip the sign of the imaginary part of the sound speed in the simulation. The resulting dispersion for the original sample ($h/d=\sqrt{3}$) is plotted in Fig.~\ref{fig5}(a). Compared with the previous dissipative case [i.e., Fig.~\ref{fig2}(b)], the real parts of the eigenfrequencies remain unchanged, while the imaginary parts have opposite signs, indicating the modes become amplifying. Due to the group velocity contrast, the modes at the $K$ valley now amplify faster during propagation, leading to a similar valley-filtering functionality. During this process, notably, the sound intensity grows as the valley polarization approaches 1 [Fig.~\ref{fig5}(b)]. Therefore, by using gain, the signal attenuation problem can be bypassed.

\end{document}